\documentclass[%
reprint,
amsmath,amssymb,
aip,
apl,
]{revtex4-2}

\usepackage{graphicx}
\usepackage{dcolumn}
\usepackage{bm}
\usepackage[mathlines]{lineno}
\usepackage{xcolor}
\usepackage[normalem]{ulem}

\usepackage{float}
\usepackage{placeins}

\usepackage[utf8]{inputenc}
\usepackage[T1]{fontenc}
\usepackage{mathptmx}
\usepackage{etoolbox}
\usepackage{physics}
\begin{document}


\title{Multitongue Frequency Fractal Dynamics in Hodgkin--Huxley Neurons Induced by Temporal Interference Stimulation}

\author{Madhurendra Mishra}
\email{madhurendramishra24@gmail.com}
\homepage{\\https://madhurendra.me}
\affiliation{Department of Physics, Sri Guru Tegh Bahadur Khalsa College, University of Delhi, New Delhi, India 110007}

\author{Mahadev Sunil Kumar}
\email{mahadevsunilkumar03@gmail.com}
\homepage{\\https://www.mahadevsunilkumar.com}
\affiliation{Indian Institute of Technology, Guwahati, Guwahati, India, 781039}

\author{Zhen Qi}
\email{zqi@wpi.edu}
\homepage{\\https://www.wpi.edu/people/staff/zqi}
\affiliation{Department of Electrical and Computer Engineering, Worcester Polytechnic Institute, Worcester, USA 01609-2280}

\author{Adarsh Ganesan}%
\email{adarsh@dubai.bits-pilani.ac.in}
\homepage{\\https://www.bits-pilani.ac.in/dr-adarsh-ganesan}
\affiliation{Department of Electrical and Electronics Engineering, Birla Institute of Technology and Science (BITS) Pilani -- Dubai Campus, Dubai International Academic City, Dubai, UAE 345055}
\affiliation{Department of Mechanical Engineering, Birla Institute of Technology and Science, Pilani - Pilani Campus, Vidya Vihar, Pilani, India 333031}

\date{\today}

\begin{abstract}
We investigate neuronal excitability in the Hodgkin--Huxley model under temporal interference (TI) stimulation in a previously unexplored Hertz-resonant regime and uncover a striking nonlinear response that we term ``multitongue frequency fractals''. Unlike single-frequency driving, which yields a smooth resonant valley, dual-frequency excitation fragments this response into a hierarchy of sharply modulated tongues whose number and structure grow with observation time, revealing clear self-similar architecture. These features emerge from transitions between non-cascaded and cascaded high-harmonic and subharmonic generation as detuning varies, and are maximised near the intrinsic ionic timescale at $\omega \approx 0.2~\mathrm{rad/ms}$. The proliferation of tongues with observation time follows a Farey-mediant hierarchy and is quantitatively consistent with a box-counting-style scaling exponent $D \approx 1.36$. These results demonstrate that TI stimulation can elicit rich, hierarchically organised frequency responses even in classical excitable membranes, revealing fractal organisation in Hodgkin--Huxley dynamics.
\end{abstract}

\maketitle

\section{Introduction}
Excitability is a universal dynamical phenomenon observed in a wide class of nonlinear systems, wherein perturbations exceeding a well-defined threshold elicit large, transient responses before the system relaxes back to a stable resting state. Originally identified in biological contexts, excitability has since emerged as a unifying concept across physics, chemistry, and engineering, providing a compact framework for understanding threshold dynamics, pulse generation, and nonlinear signal processing \cite{Izhikevich2007,Pikovsky2001}.

From a dynamical systems perspective, excitability typically arises near bifurcation points, such as saddle-node or Hopf bifurcations, where slow-fast time-scale separation plays a crucial role. In such regimes, trajectories may undergo large excursions in phase space despite the absence of sustained oscillations, a behaviour that has been extensively studied in electronic circuits, optical cavities, chemical reactions, and mechanical resonators \cite{FitzHugh1961,Meron1992}. These physical realisations highlight that excitability is not restricted to living systems but is instead a generic consequence of nonlinear feedback and thresholded instability.

The canonical theoretical description of excitability in neuroscience is provided by the Hodgkin--Huxley (HH) model, which describes the action potential as a nonlinear dynamical response of the neuronal membrane driven by voltage-dependent ion channel kinetics \cite{Hodgkin1952,nelson1998hodgkin,yuan2016theoretical}. Beyond its biological relevance, the HH model has profoundly influenced the study of excitable media in physics, inspiring reduced descriptions such as the FitzHugh--Nagumo model and motivating experimental investigations in synthetic and engineered systems \cite{FitzHugh1961,levi2018digital,rutherford2020analog, kumar2016design, giannari2022model}. These developments established excitability as a paradigmatic nonlinear phenomenon characterised by robustness, reproducibility, and sensitivity to weak stimuli.

In 2017, a significant conceptual advance in the control of excitable systems occurred with the proposal of temporal interference (TI) stimulation by Grossman et al. \cite{Grossman2017}. The central idea relies on the superposition of two high-frequency electric fields with slightly detuned frequencies, producing a low-frequency envelope capable of selectively modulating neural excitability without directly stimulating superficial tissue. Although individual carrier frequencies lie beyond the neuronal firing bandwidth, their interference generates an effective modulation that can cross excitability thresholds in targeted regions.

Subsequent experimental and theoretical studies demonstrated that temporal interference can induce spiking, modulate firing rates, and achieve spatially selective deep-brain stimulation, thereby revealing a new route for noninvasive control of excitable dynamics \cite{Mirzakhalili2020,Esmaeilpour2021}. From a nonlinear dynamics standpoint, TI stimulation exploits the intrinsic threshold structure of excitable systems, effectively reshaping the input landscape without altering the underlying resting state.

Despite growing experimental support, the mechanisms underlying temporal interference stimulation were investigated only recently based on Hodgkin--Huxley dynamics \cite{plovie2022influence,plovie2024contribution,plovie2025nonlinearities}. These studies showed that, for kilohertz-range carriers, high-frequency sinusoidal stimulation drives neurons into a stable subthreshold periodic state maintained by a cycle-averaged balance between inward sodium and outward potassium currents. This cycle-averaged description relies on a clean separation between the carrier period and the gating time constants of the sodium and potassium channels, so that the fast carrier can be adiabatically eliminated and only the slow envelope controls excitability. The carriers considered here instead place the drive frequency close to these intrinsic ionic timescales, so that adiabatic elimination no longer applies and the gating variables respond directly to individual carrier cycles rather than to their time average. Our results therefore extend Refs.~\cite{plovie2022influence,plovie2024contribution,plovie2025nonlinearities} to this non-adiabatic regime, in which the interplay between the carrier period and channel kinetics itself generates the harmonic and subharmonic resonances underlying the multitongue structure, rather than contradicting the cycle-averaged mechanism identified at kilohertz frequencies, which we expect to be recovered as a limiting case at sufficiently high carrier frequency.

Although temporal interference stimulation is generally limited to frequencies in the kilohertz range, we chose a frequency in the hertz range allowing resonant excitation of Hodgkin--Huxley dynamics similar to \cite{parmananda2001resonance,parmananda2002resonant}. Because two weakly detuned carriers constitute a quasiperiodic drive, our setup is also related to prior studies of quasiperiodically forced excitable membranes, which have so far been restricted to reduced models such as the Morris--Lecar neuron \cite{lim_kim_kim_2009}; here we extend this line of inquiry to the full Hodgkin--Huxley dynamics. By doing so, we uncover a previously unreported synchronisation regime in which temporal interference stimulation induces a recursively organised hierarchy of resonance tongues. As we show below, the resulting multitongue frequency fractals are governed by Farey-sequence arithmetic and exhibit self-similar proliferation with increasing observation time.

\begin{figure*}
    \centering
    \includegraphics[width=0.9\linewidth]{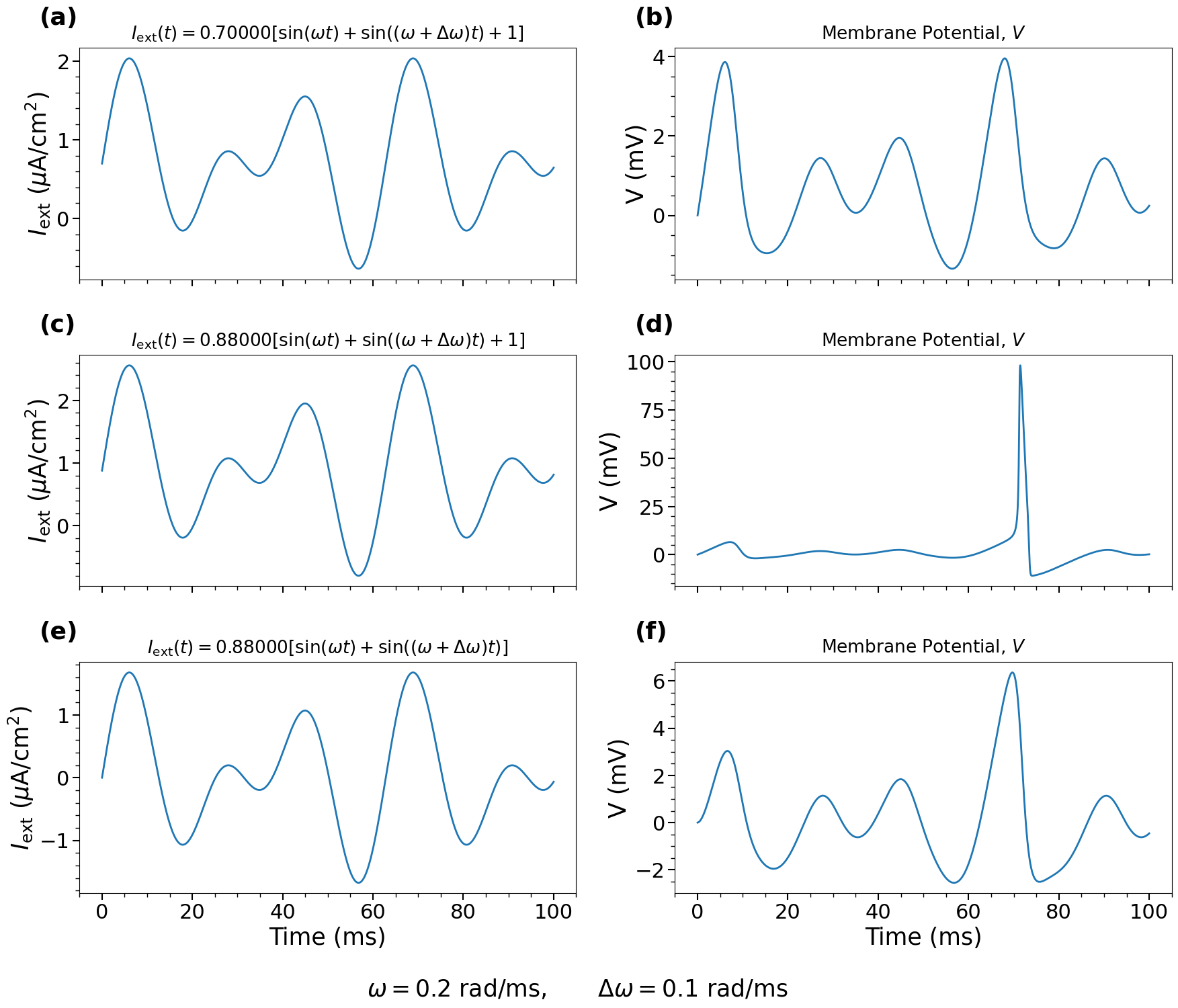}
    \caption{Time-domain responses of the Hodgkin--Huxley neuron to dual-frequency stimulation at $\omega=0.2~\mathrm{rad/ms}$ and $\Delta\omega=0.1~\mathrm{rad/ms}$. Panels (a,b) show the subthreshold response with the DC offset, (c,d) the spiking response with the DC offset, and (e,f) the response without the DC offset.}
    \label{fig0}
\end{figure*}

\begin{figure}
    \centering
    \includegraphics[width=0.95\linewidth]{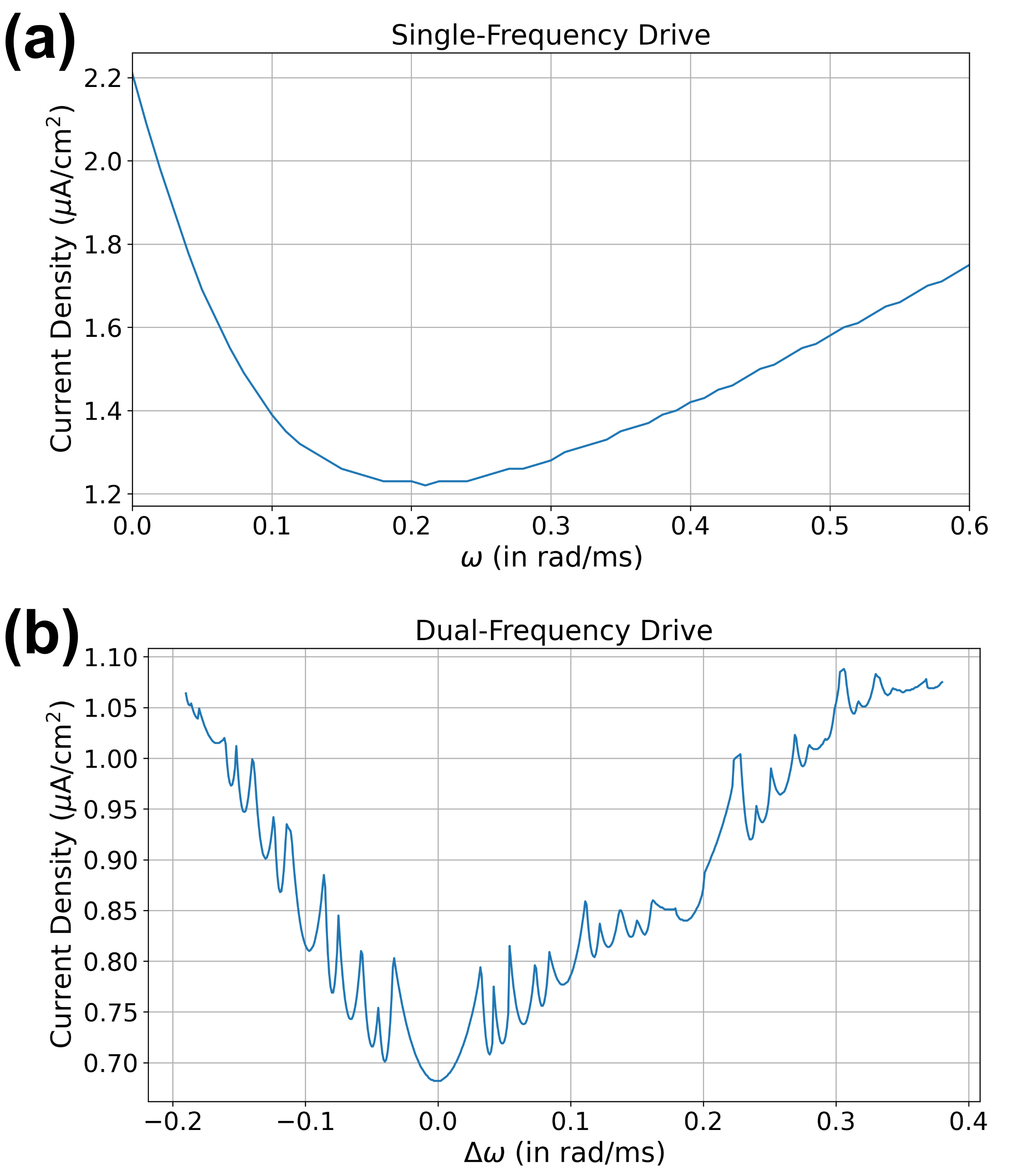}
    \caption{Stimulation threshold amplitude $J_y^*$ under (a) single-frequency and (b) dual-frequency excitation. In (a), $J_y^*$ reaches its minimum near $\omega \approx 0.2~\mathrm{rad/ms}$. In (b), for $\omega = 0.2~\mathrm{rad/ms}$, the dual-frequency drive produces a broader response with multiple minima in $J_y^*$, giving rise to the observed ``multitongue fractal'' structure.}
    \label{fig1}
\end{figure}

\begin{figure*}
    \centering
    \includegraphics[width=0.9\linewidth]{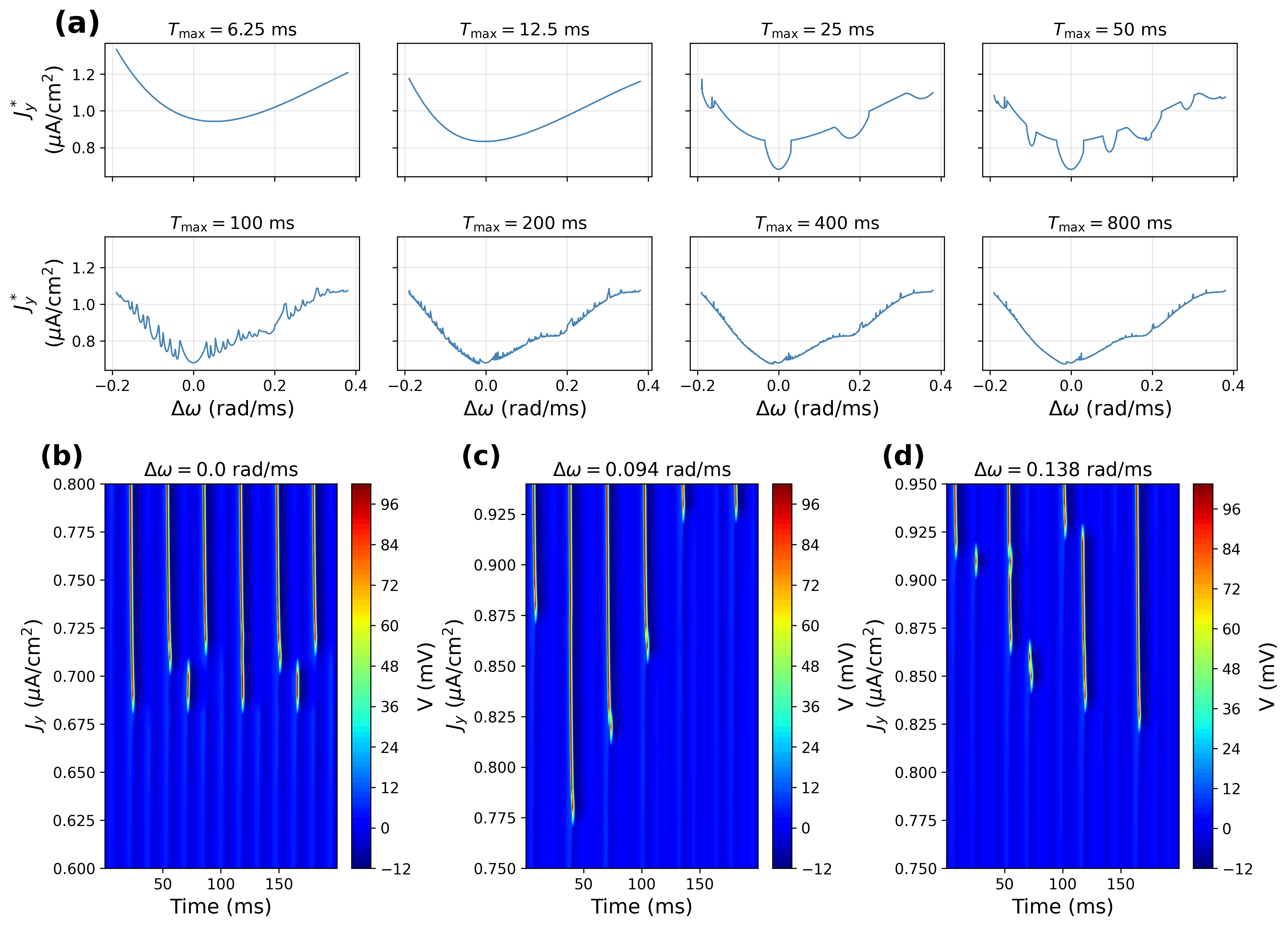}
    \caption{(a) Stimulation threshold amplitude $J_y^*$ under dual-frequency excitation for different maximum integration times $T_{\max}$, showing the emergence of tongues within valleys as fractal-like structures develop with increasing observation time. As $T_{\max}$ increases, additional tongue-like substructures appear within the primary valleys, producing a self-similar, fractal-like frequency response. Panels (b)–(d) illustrate three representative dynamical behaviours underlying this structure: (b) non-cascaded high-harmonic excitation at $\Delta\omega = 0$, (c) cascaded high-harmonic excitation at $\Delta\omega \approx 0.094~\mathrm{rad/ms}$, and (d) cascaded subharmonic excitation at $\Delta\omega \approx 0.138~\mathrm{rad/ms}$. The coexistence and transitions among these distinct response regimes collectively generate the hierarchical ``frequency fractals'' observed in (a).}
    \label{fig2}
\end{figure*}

Foundational studies established that periodically forced Hodgkin--Huxley neurons exhibit harmonic and subharmonic synchronisation, quasiperiodicity, and chaos under sinusoidal stimulation \cite{aihara_periodic_1984,degn_forced_1987}. Subsequent work showed that these responses organise into Arnold tongues corresponding to rational ($p:q$) locking states and are structured by codimension-two bifurcations \cite{che_phase-locking_2009,lee_bifurcation_2006}. This mode-locking phenomenology, comprising proliferating rational frequency-locking regions, Farey-ordered tongues, and devil's-staircase organisation, has been documented extensively in periodically and quasiperiodically forced nonlinear oscillators since the 1980s \cite{guevara_glass_1982,perez_glass_1982,grebogi_ott_yorke_1985}. Beyond Hodgkin--Huxley dynamics, synchronisation hierarchies in nonlinear oscillators were shown to possess an arithmetic organisation governed by Farey sequences, devil's staircases, and mediant insertion rules linking neighbouring rational states \cite{allen_arithmetic_1983,abhay_high-order_2025,felicio_arnold_2015}. Collectively, these studies established synchronisation as both a geometric and arithmetic phenomenon, but almost exclusively as a property of a single forcing frequency evaluated in the asymptotic long-time limit.

Despite these advances, previous studies primarily characterised synchronisation through localised structures such as Arnold tongues, bifurcation curves, Poincaré sections, return maps, and firing-rate statistics \cite{aihara_periodic_1984,borkowski_response_2009,borkowski_bistability_2011}. In this classical picture, the tongue structure is a fixed geometric object in the (frequency, amplitude) plane evaluated once transients have decayed: only a single forcing frequency is considered, and the observation window itself is not treated as a control parameter. Here we show that temporal interference stimulation fundamentally alters this picture along both axes. Because the two carriers are only weakly detuned, their interference generates cascades of high- and sub-harmonic mode-locked states that a single tone at the same amplitude does not produce. More importantly, we show that the observation window $T_{\max}$ is itself an organising parameter rather than a passive averaging window: instead of converging to a fixed asymptotic tongue structure, the excitation landscape recursively fragments as $T_{\max}$ grows, with new tongues emerging between existing ones in a manner consistent with Farey-mediant ordering. The mere existence of rational locking is indeed generic to forced nonlinear oscillators; the phenomenon reported here is instead this $T_{\max}$-driven proliferation itself, a hierarchical, self-similar resonance architecture in the time domain that static Arnold-tongue descriptions do not capture. In this framework, Farey sequences become the organising mechanism responsible for the proliferation of resonance structures across scales.

\section{Mathematical Model}

The HH model describes the electrical activity of a neuron by representing the cell membrane as an electrical circuit composed of a capacitor and nonlinear ionic conductors \cite{yuan2016theoretical, johnson2017spike}. The membrane dynamics are governed by the following set of coupled differential equations:

\begin{equation}
\label{eq:hh}
\begin{split}
C_m \frac{dV}{dt}
&=
I_{\text{ext}}
-
\Big[
\bar{g}_{\mathrm{Na}} m^3 h (V - V_{\mathrm{Na}})
\\
&\quad
+
\bar{g}_{\mathrm{K}} n^4 (V - V_{\mathrm{K}})
+
g_L (V - V_L)
\Big]
\end{split}
\end{equation}

Here, $V$ denotes the membrane potential, $C_m$ is the membrane capacitance per unit area, $I_{\text{ext}}$ represents the externally applied current, $\bar{g}_{\mathrm{Na}}$ is the sodium conductance, $V_{\mathrm{Na}}$ is the sodium equilibrium potential, $\bar{g}_{\mathrm{K}}$ is the potassium conductance, $V_{\mathrm{K}}$ is the potassium equilibrium potential, $g_L$ is the leak conductance and $V_L$ is the leak reversal potential.

The dynamics of the gating variables $m$, $h$, and $n$ are governed by first-order kinetics:
\begin{equation}
\begin{aligned}
\frac{dm}{dt}
=
\varphi \left[ \alpha_m(V)(1 - m) - \beta_m(V)m \right]
\\
\frac{dh}{dt}
=
\varphi \left[ \alpha_h(V)(1 - h) - \beta_h(V)h \right]
\\
\frac{dn}{dt}
=
\varphi \left[ \alpha_n(V)(1 - n) - \beta_n(V)n \right]
\end{aligned}
\end{equation}
Each gating variable represents the probability that a specific channel gate is open. The functions $\alpha_x(V)$ and $\beta_x(V)$ denote voltage-dependent opening and closing rates, respectively.

The rate constants are defined as follows:

\begin{equation*}
\begin{aligned}
\alpha_m(V) &= \frac{0.1(25 - V)}{\exp\!\left(\frac{25 - V}{10}\right) - 1} \\
\beta_m(V)  &= 4 \exp\!\left(-\frac{V}{18}\right) \\
\alpha_h(V) &= 0.07 \exp\!\left(-\frac{V}{20}\right) \\
\beta_h(V)  &= \frac{1}{\exp\!\left(\frac{-V + 30}{10}\right) + 1} \\
\alpha_n(V) &= \frac{0.01(10 - V)}{\exp\!\left(\frac{10 - V}{10}\right) - 1} \\
\beta_n(V)  &= 0.125 \exp\!\left(-\frac{V}{80}\right)
\end{aligned}
\end{equation*}

The removable singularities in $\alpha_m(V)$ and $\alpha_n(V)$ at $V=25~\mathrm{mV}$ and $V=10~\mathrm{mV}$, respectively, are handled numerically by evaluating their limiting values, giving $\alpha_m(25)=1$ and $\alpha_n(10)=0.1$.

These expressions are empirically derived to fit voltage-clamp experimental data and capture the nonlinear voltage dependence of ion channel kinetics \cite{yuan2016theoretical}.

The parameter $\varphi$ accounts for temperature-dependent changes in channel kinetics and is defined as $\varphi = 3^{\frac{T - 6.3}{10}}$.
Here, the temperature $T$ is set to $6.3^\circ\mathrm{C}$, yielding $\varphi=1$, corresponding to the reference temperature of the classical Hodgkin--Huxley formulation. We chose $6.3^\circ\mathrm{C}$ deliberately because
we did not consider the effect of temperature on the electrical
activity of neurons. However, temperature dependence, spatial effects, and stochasticity can be incorporated in future work to assess the robustness of the reported phenomena under more physiologically realistic conditions.

\section{Numerical Simulation Setup}

We numerically investigate neuronal excitability under TI stimulation using a single-compartment HH neuron model. The parameter values are set as $C_m = 1~\mu\mathrm{F/cm}^2$, $\bar{g}_{\mathrm{K}} = 36~\mathrm{mS/cm}^2$, $g_L = 0.3~\mathrm{mS/cm}^2$ and $\bar{g}_{\mathrm{Na}} = 120~\mathrm{mS/cm}^2$, with potentials $V_{\mathrm{Na}} = 115~\mathrm{mV}$, $V_{\mathrm{K}} = -12~\mathrm{mV}$ and $V_L = 10.59~\mathrm{mV}$. To set the resting membrane potential to zero, the membrane voltage is shifted by $65~\mathrm{mV}$. The initial conditions are set as $
V(0) = 0, m(0) = 0.053, h(0) = 0.596, n(0) = 0.317$. These values correspond to the steady-state resting conditions of the HH neuron model.

The temporal interference stimulation current $I_{\mathrm{ext}}$ is defined as
\begin{equation}
\begin{split}
I_{\mathrm{ext}}(t)
&= J_y \Big[
\sin(\omega t)
+ \sin\!\big((\omega + \Delta\omega)t\big)
+ 1
\Big]
\end{split}
\label{eq:iext}
\end{equation}
where $J_y$ is the stimulation amplitude, $\omega$ is the carrier angular frequency, and $\Delta\omega$ is the detuning. Throughout the simulations, the carrier frequency is fixed at $\omega = 0.2~\mathrm{rad/ms}$, while $\Delta\omega$ is systematically varied, enforcing a clear separation of timescales which is required for effective TI stimulation. Owing to the intrinsic low-pass filtering properties of neuronal membranes, the neuron is largely insensitive to the high-frequency carrier and responds primarily to the slow envelope modulation.

For the numerical parameter sweep, the detuning was sampled on a uniform grid with spacing $\delta(\Delta\omega)=0.001~\mathrm{rad/ms}$, yielding 571 detuning points over the range $-0.190\leq\Delta\omega\leq0.380~\mathrm{rad/ms}$. For each detuning value, the threshold amplitude $J_y^*$ was obtained by binary search with a step size of $0.001~\mu\mathrm{A/cm^2}$. The numerical integration time step was $dt=0.0002~\mathrm{ms}$.

Spike initiation is defined by a membrane potential crossing a fixed threshold, $V(t) > V_{\mathrm{th}} = 50~\mathrm{mV}$.
The HH equations are integrated forward in time, and the minimum stimulation amplitude $J_y$ required to elicit at least one spike is determined using a binary search algorithm. This procedure assumes monotonicity of spike initiation with respect to stimulation amplitude and yields the firing threshold function $J_y^*$.

We explicitly verified the monotonicity of spike initiation with respect to $J_y$ at representative detunings spanning the resonance regions. In all tested cases, the response transitioned from no spiking to spiking without subsequent loss of spiking as $J_y$ increased, supporting the use of the binary-search procedure for determining $J_y^*$.

\begin{figure}
	\centering
	\includegraphics[width=0.9\linewidth]{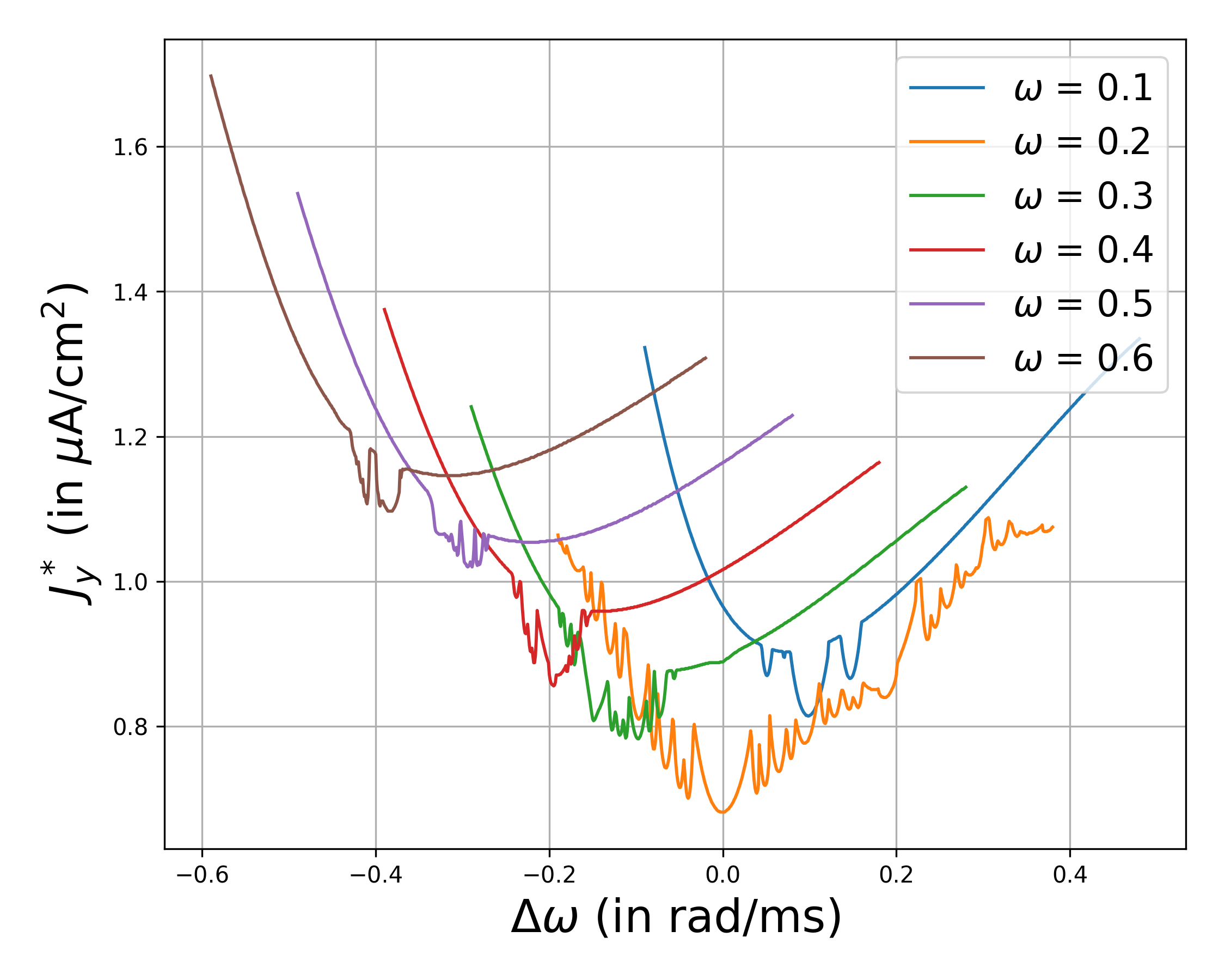}
    \caption{Stimulation threshold amplitude $J_y^*$ under dual-frequency excitation for different primary drive frequencies $\omega$ as a function of detuning $\Delta\omega$. A fractal-like modulation of the response is most prominently observed near the resonant condition at $\omega = 0.2~\mathrm{rad/ms}$, where the system displays the widest spread and highest complexity of oscillatory structures. As $\omega$ increases away from resonance, these frequency-fractal features become confined to a progressively narrower detuning range and their intensity is significantly reduced, leading to smoother response curves.}
	\label{fig3}
\end{figure}

\begin{figure*}
	\centering
	\includegraphics[width=\linewidth]{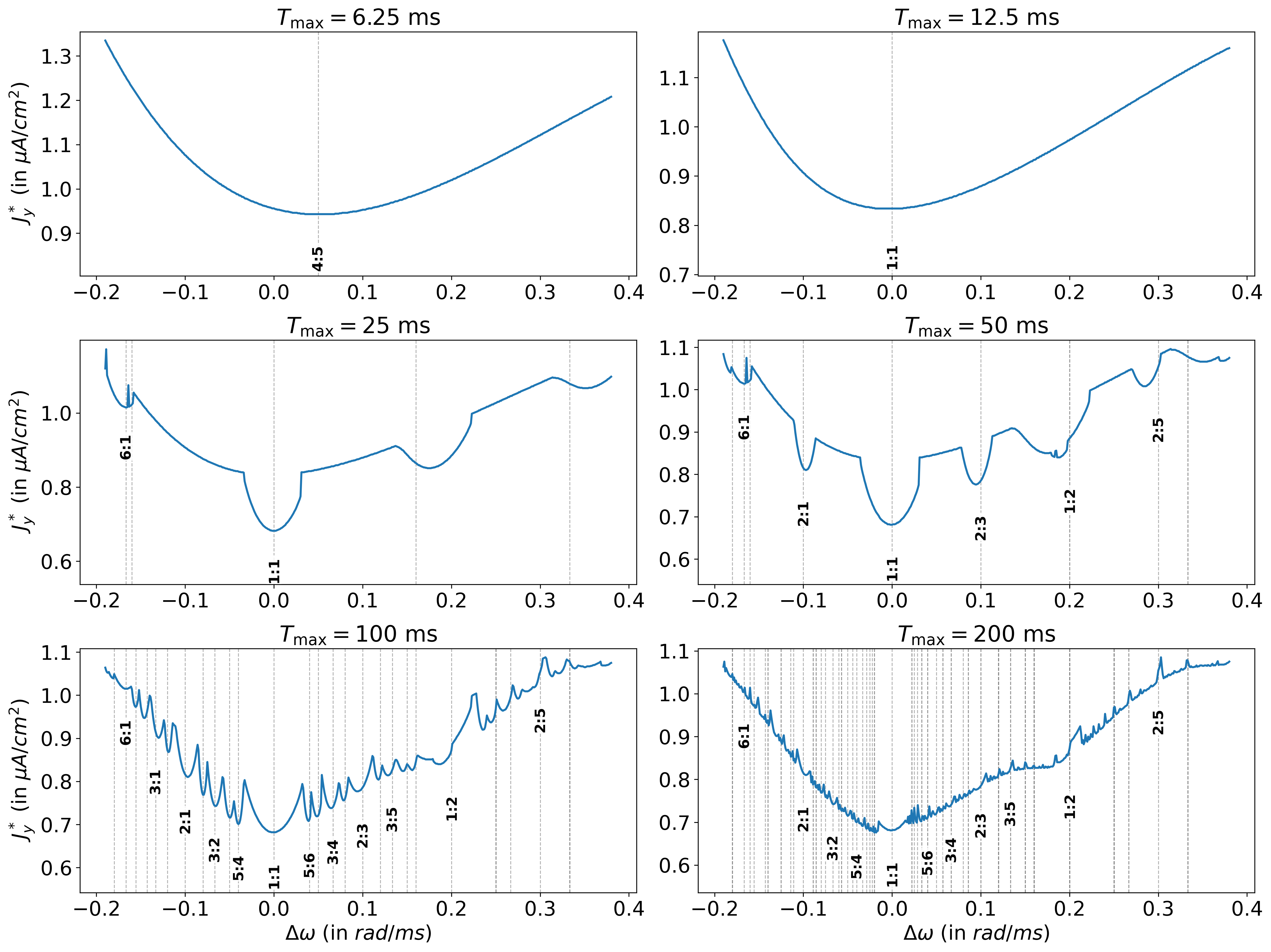}
	\caption{
		Evolution of multitongue frequency fractals with increasing observation time $T_{\max}$. Dashed vertical lines indicate local minima in the excitation threshold $J_y^*$ and are labelled by the corresponding rational locking ratios $p:q$ extracted from the dominant spectral response. As $T_{\max}$ increases, additional synchronisation tongues emerge between previously existing resonances in a manner consistent with Farey-mediant ordering. Principal resonances such as $1:1$, $1:2$, and $2:1$ appear first, while higher-order resonances progressively fill the intervals between them, producing a recursive refinement of the resonance landscape. The resulting hierarchy demonstrates that multitongue proliferation follows a Farey-tree organisation and provides direct evidence that the observed frequency fractals possess an underlying arithmetic structure.
	}
	\label{fig:farey}
\end{figure*}

\section{Results and Discussion}

Figure~\ref{fig0} shows the time-domain response of the Hodgkin--Huxley neuron under dual-frequency stimulation with $\omega=0.2~\mathrm{rad/ms}$ and $\Delta\omega=0.1~\mathrm{rad/ms}$. The responses in panels (a) and (b) remain subthreshold at lower stimulation amplitudes, whereas a further increase in $J_y$ produces clear action potentials in panels (c) and (d) in the presence of the DC offset. By contrast, panels (e) and (f) remain subthreshold at the same stimulation amplitude when the DC component is removed. Despite a higher threshold, the multitongue frequency-fractal structure persists without the DC offset.

Figure~\ref{fig1} establishes the fundamental contrast between single- and dual-frequency stimulation of the Hodgkin--Huxley membrane. Under single-frequency driving, the stimulation threshold amplitude $J_y^*$ varies smoothly with frequency and exhibits a shallow minimum near the intrinsic resonance. However, when a second tone is introduced, this simple valley fragments into a series of sharp lobes and dips, which we term multitongue frequency fractals. The emergence of these distinct modulation tongues demonstrates strong nonlinear frequency mixing and indicates that the membrane acts as a nonlinear envelope detector that is highly sensitive to the slow beat component generated by temporal interference.

The hierarchical structure of these modulation tongues is revealed in Figure~\ref{fig2}. As the observation window $T_{\max}$ is increased, smooth valleys split into progressively finer tongue-like features, resulting in a self-similar, fractal-like frequency landscape.
The membrane-potential dynamics shown in the colour maps of Figure~\ref{fig2}(b)--(d) provide the dynamical origin of this structure. In these panels, the horizontal axis denotes time, the vertical axis denotes the stimulation amplitude $J_y$, and the colour represents the instantaneous membrane potential $V(t)$. Thus, the high-voltage bands directly identify action-potential excursions and their evolution with increasing $J_y$, allowing the distinct harmonic and subharmonic response regimes to be distinguished across the full amplitude range. At $\Delta\omega = 0$, the response corresponds to non-cascaded high-harmonic generation wherein all harmonics are excited simultaneously above $J_y\approx 0.68~\mu\mathrm{A/cm^2}$; at intermediate detuning ($\Delta\omega \approx 0.094~\mathrm{rad/ms}$), the system exhibits cascaded high-harmonic generation wherein the fundamental tone is excited first and the higher-order harmonics are subsequently excited at higher $J_y$ values. At larger detuning ($\Delta\omega \approx 0.138~\mathrm{rad/ms}$), cascaded subharmonic excitation is observed wherein the subharmonics are generated as $J_y$ is increased. The coexistence of these qualitatively distinct regimes, along with sharp transitions among them as detuning varies, generates the observed ``multitongue frequency fractals''. These harmonic and subharmonic responses are conceptually analogous to the $N{:}1$ subharmonic and superharmonic locking rhythms documented in classical periodically forced Hodgkin--Huxley studies \cite{aihara_periodic_1984,che_phase-locking_2009}. To verify this, we performed spectral analysis of the subthreshold voltage traces (no action potentials present), which confirms the fundamental carrier frequency to be $\omega = 0.2~\mathrm{rad/ms}$ as stated: at $\Delta\omega = 0$, the fundamental component at $\omega$ is present but subdominant, while the response is instead dominated by its second harmonic at $2\omega$, consistent with the non-cascaded high-harmonic generation described above. Note that this harmonic content of the neuronal response is a distinct notion from the rational $p{:}q$ labels used in Figure~\ref{fig:farey}, which classify tongues by the commensurability ratio between the two carrier frequencies $\omega$ and $\omega+\Delta\omega$ rather than by the response spectrum; establishing the precise correspondence between the two classifications is a natural extension of the present analysis.

Figure~\ref{fig3} demonstrates that this phenomenon is resonantly organised. When the primary driving frequency is tuned near $\omega = 0.2~\mathrm{rad/ms}$, the tongues are the most numerous, extend over the widest detuning range, and display the highest amplitude, indicating maximal nonlinear mixing. As the carrier frequency moves away from resonance, the fractal structures compress and the response gradually becomes smoother. This shows that the fractal behaviour is not incidental, but is resonantly amplified when the temporal interference envelope matches the intrinsic ionic timescales.

To determine whether the observed multitongue structures possess an underlying arithmetic organisation in the form of a Farey sequence, each local minimum identified in Figure~\ref{fig:farey} was assigned a rational locking ratio $p/q$ obtained from the dominant spectral response. The newly resolved tongues systematically appear between neighbouring resonances according to the Farey mediant rule
\begin{equation}
\frac{p_1}{q_1},
\frac{p_2}{q_2}
\rightarrow
\frac{p_1+p_2}{q_1+q_2}.
\end{equation}

For example, the neighbouring resonances $1:1$ and $1:2$ generate the mediant resonance $2:3$, while the neighbouring resonances $2:3$ and $1:2$ generate $3:5$. Higher-order resonances such as $5:8$ emerge through repeated mediant insertion. Similar relationships are observed throughout the hierarchy as $T_{\max}$ increases. Consequently, the proliferation of multitongues follows a recursive Farey-tree construction rather than an arbitrary refinement process.

This interpretation is further supported by the ordering of the detected resonances in detuning space. Adjacent synchronisation states satisfy the Farey-neighbour condition $|p_1 q_2 - p_2 q_1| = 1$, which is characteristic of consecutive fractions in a Farey sequence. The observed ordering shown in Figure~\ref{fig:farey} is consistent with successive generations of Farey refinement, with low-order resonances appearing at short observation times. As additional observation time reveals progressively weaker resonances, higher-order mediants become visible, generating increasingly dense synchronisation structures near the principal $1:1$ resonance. The resulting resonance landscape therefore combines geometric self-similarity with arithmetic self-organisation. In this sense, the multitongue frequency fractals can be interpreted as a dynamical realisation of a Farey hierarchy embedded within the temporal interference response of the Hodgkin--Huxley neuron.

\begin{figure}
    \centering
    \includegraphics[width=0.9\linewidth]{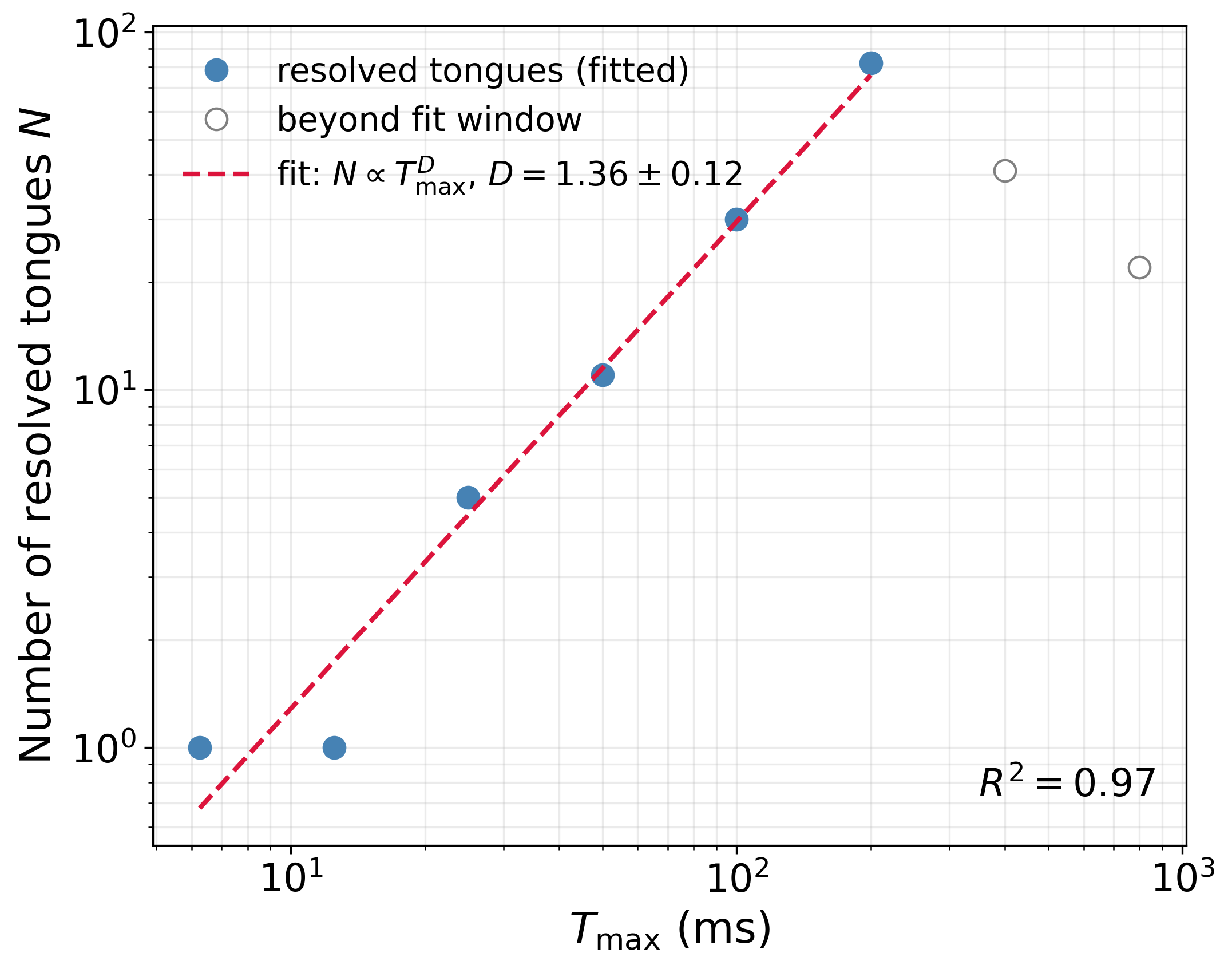}
    \caption{Box-counting analysis of multitongue proliferation. The number of resolved tongues $N$ (local minima of $J_y^*$, using a fixed peak-detection threshold) is counted at each $T_{\max}$ from the data underlying Figure~\ref{fig2}(a). Over $T_{\max} = 6.25$--$200~\mathrm{ms}$ (filled circles), $N$ grows as a power law, $N \propto T_{\max}^{D}$ with $D = 1.36 \pm 0.12$ ($R^2 = 0.97$). Beyond $T_{\max} \approx 200~\mathrm{ms}$ (open circles), the tongue count declines as neighbouring high-order resonances merge, bounding the scaling window.}
    \label{fig:boxcount}
\end{figure}

\subsection{Box-Counting Analysis of Multitongue Fractal Proliferation}

To quantify the observation-time dependence of the multitongue frequency fractals, we examined the scaling of the number of resolved resonance tongues with $T_{\max}$.\\

\textbf{\textit{Scaling rationale}}--- Figure~\ref{fig2}(a) shows that increasing $T_{\max}$ reveals progressively finer resonance tongues within the primary valleys of $J_y^*(\Delta\omega)$. To quantify this, we analysed the number of resolved tongues as a function of observation time. Motivated by the conventional box-counting relation $D = \lim_{\varepsilon \to 0} \log N(\varepsilon)/\log(1/\varepsilon)$, we examined whether the tongue count follows a power law, $N(T_{\max}) \propto T_{\max}^{D}$. Here, $D$ is an empirical scaling exponent characterising tongue proliferation over the observation time range considered, rather than a formally established fractal dimension.\\

\textbf{\textit{Tongue counting procedure}}--- For each of the eight observation windows, $T_{\max} = 6.25$, $12.5$, $25$, $50$, $100$, $200$, $400$, and $800~\mathrm{ms}$, we counted the local minima of $J_y^*(\Delta\omega)$ across the detuning interval $\Delta\omega \in [-0.2, 0.38]~\mathrm{rad/ms}$. The minima were identified using SciPy's \texttt{find\_peaks} applied to $-J_y^*(\Delta\omega)$, with a prominence threshold of $0.005~\mu\mathrm{A/cm^2}$ and a minimum separation of two samples. These parameters match those used to identify resonance tongues in the preceding analysis. The resulting counts, denoted by $N(T_{\max})$, are summarised in Table~\ref{tab:tongue-counts}.\\

\begin{table}[htbp]
\centering
\caption{\bf Number of resolved tongues $N(T_{\max})$}
\begin{tabular}{ccccccccc}
\hline
$T_{\max}$ (ms) & 6.25 & 12.5 & 25 & 50 & 100 & 200 & 400 & 800 \\
\hline
$N(T_{\max})$ & 1 & 1 & 5 & 11 & 30 & 82 & 41 & 22 \\
\hline
\end{tabular}
\label{tab:tongue-counts}
\end{table}

\textbf{\textit{Scaling analysis}}--- The tongue count increases from $N=1$ at $T_{\max}=6.25~\mathrm{ms}$ to $N=82$ at $T_{\max}=200~\mathrm{ms}$, before declining at longer observation times. We therefore fitted the power law $N(T_{\max}) \propto T_{\max}^{D}$ over the monotonically increasing range $6.25 \leq T_{\max} \leq 200~\mathrm{ms}$. Linear regression of $\log N$ against $\log T_{\max}$ yields $D=1.361\pm0.119$, with $R^2=0.970$. The fitted interval exhibits a clear power-law trend, whereas the counts at $400$ and $800~\mathrm{ms}$ fall to $41$ and $22$, respectively. These points are excluded from the fit. Their decline indicates that the detected tongue count does not increase monotonically across the full observation time range, potentially reflecting the merging of neighbouring features or limitations of the peak detection procedure. Figure~\ref{fig:boxcount} distinguishes the fitted points from these longer observation-time results.\\

\textbf{\textit{Sensitivity to peak detection}}--- Because the measured tongue count depends on the criterion used to distinguish individual minima, we assessed the sensitivity of the fitted exponent to the prominence threshold and minimum peak separation. We repeated the regression over the same observation time interval, $6.25 \leq T_{\max} \leq 200~\mathrm{ms}$, varying the prominence threshold from $0.002$ to $0.020~\mu\mathrm{A/cm^2}$ and the minimum peak separation from one to five samples. Table~\ref{tab:sensitivity} reports the resulting exponents and coefficients of determination.\\

\begin{table}[htbp]
\centering
\caption{\bf Sensitivity of the fitted exponent $D$ to the peak-detection threshold}
\begin{tabular}{cccc}
\hline
Prominence ($\mu$A/cm$^2$) & Distance (samples) & $D$ & $R^2$ \\
\hline
0.002 & 2 & 1.426 & 0.969 \\
0.003 & 2 & 1.397 & 0.970 \\
0.005 & 2 (baseline) & 1.361 & 0.970 \\
0.008 & 2 & 1.277 & 0.968 \\
0.010 & 2 & 1.234 & 0.964 \\
0.020 & 2 & 0.997 & 0.903 \\
0.005 & 1 & 1.361 & 0.970 \\
0.005 & 3 & 1.329 & 0.970 \\
0.005 & 5 & 1.272 & 0.967 \\
\hline
\end{tabular}
\label{tab:sensitivity}
\end{table}

The fitted exponent varies from approximately $1.00$ to $1.43$ across the tested detection settings. At the baseline prominence of $0.005~\mu\mathrm{A/cm^2}$, increasing the minimum peak separation from two to five samples reduces the exponent from $1.361$ to $1.272$. Increasing the prominence threshold to $0.020~\mu\mathrm{A/cm^2}$ yields $D=0.997$ and reduces the fit quality to $R^2=0.903$, consistent with the suppression of weaker minima. The persistence of a strong power-law fit across most tested settings supports the presence of systematic tongue proliferation over the fitted observation time interval. However, the dependence of $D$ on the detection parameters, together with the decline in tongue counts at longer observation times, warrants interpreting $D$ as an empirical scaling exponent rather than definitive evidence of a fractal dimension.

\subsection{Circle-Map Interpretation of the Multitongue Fractal Response}

In this section, we develop an analytical reduction for the two-frequency
HH system. Starting from
the applied temporal-interference current, we identify the slow relative
phase of the two tones, show how nonlinear mixing generates a
difference-frequency contribution, and construct an event-based return
map from successive spike times. Under a first-harmonic approximation,
this return map reduces to the standard sine circle map, which provides
the natural framework for defining the winding number, rational mode
locking, Arnold tongues, and Farey organisation.

The HH system (Eq.~\eqref{eq:hh}) can be written compactly as

\begin{equation}
\dot{\mathbf{X}} = \mathbf{F}(\mathbf{X}) + \mathbf{b}I_{\mathrm{ext}},
\qquad
\mathbf{b} = \begin{pmatrix}
\dfrac{1}{C_m} \\ 0 \\ 0 \\ 0
\end{pmatrix}.
\end{equation}

where the state vector $\mathbf{X} = (V,m,h,n)^{T}$.

Equation~\eqref{eq:iext} can be rewritten as

\begin{equation}
    \begin{aligned}
        I_{\mathrm{ext}}(t) &= J_y\left\{1 + 2\cos\left(\frac{\Delta\omega t}{2}\right)
        \sin(\omega_c t)\right\},\\
        \omega_c &= \omega + \frac{\Delta\omega}{2}.
    \end{aligned}
\end{equation}

The carrier-envelope form identifies a relatively fast carrier at
\(\omega_{c}\) modulated by a slowly varying interference envelope. The
natural slow phase is the relative phase of the two tones, $\psi(t) = \Delta\omega t,\quad\quad\dot{\psi} = \Delta\omega$. Constructive/destructive interference repeats whenever the relative
phase advances by \(2\pi\), so the beat period is $T_{b} = \frac{2\pi}{|\Delta\omega|}$.

There is an apparent factor-of-two issue if one looks only at the signed
factor \(\cos(\Delta\omega t/2)\), whose algebraic period is
\(4\pi/|\Delta\omega|\). However, changing the sign of this factor is
equivalent to shifting the fast carrier by \(\pi\). The physically
relevant interference pattern therefore repeats after a relative phase
change of \(2\pi\), giving the beat-period expression above. This explicitly identifies the slow phase variable underlying the
envelope response.

The carrier-envelope identity alone shows amplitude modulation. A
dynamical explanation for sensitivity to the slow difference frequency
follows from nonlinear mixing in the HH equations. Let \(\mathbf{X}_{r}\) be the resting state and write $\mathbf{X} = \mathbf{X}_{r} + \mathbf{x}$.

Expand the nonlinear vector field about \(\mathbf{X}_{r}\):

\begin{equation}
    \begin{aligned}
        \dot{\mathbf{x}} &= A\mathbf{x}
        + \frac{1}{2}Q(\mathbf{x},\mathbf{x})
        + \frac{1}{6}C(\mathbf{x},\mathbf{x},\mathbf{x})
        + \cdots \\[4pt]
        &\quad + J_y\mathbf{b}
        \left[\sin(\omega t) + \sin\bigl((\omega + \Delta\omega)t\bigr)\right].
    \end{aligned}
\end{equation}

where $A = D\mathbf{F}\left( \mathbf{X}_{r} \right)$ is the Jacobian and \(Q\) and \(C\) denote quadratic and cubic terms of
the Taylor expansion.


At first order, the response contains the two drive frequencies,

\begin{equation}
\begin{aligned}
\mathbf{x}_{1}(t) &\simeq \operatorname{Re}\Bigl[
\mathbf{H}_{1}e^{i\omega t} \\
&\qquad + \mathbf{H}_{2}e^{i(\omega+\Delta\omega)t}\Bigr],
\end{aligned}
\label{eq:linear-two-tone-response}
\end{equation}

where, schematically,

\begin{equation}
\begin{aligned}
\mathbf{H}_{j} &= (i\omega_j I-A)^{-1}\mathbf{b}\,\widetilde{J}_{y},\\
\omega_{1} &= \omega,\qquad \omega_{2}=\omega+\Delta\omega.
\end{aligned}
\label{eq:linear-susceptibility}
\end{equation}

Substituting the first-order response in Eq.~\eqref{eq:linear-two-tone-response} into the quadratic term produces
cross-products between the two frequencies. In complex notation, $e^{i(\omega + \Delta\omega)t}e^{- i\omega t} = e^{i\Delta\omega t}$, so a contribution at the difference frequency \(\Delta\omega\)
necessarily appears whenever the relevant quadratic coupling coefficient
is nonzero. Thus the averaged nonlinear dynamics has the generic form

\begin{equation}
\begin{aligned}
\bigl\langle Q(\mathbf{x}_{1},\mathbf{x}_{1})\bigr\rangle_{\mathrm{fast}}
&\sim B_{0} + B_{1}\cos(\Delta\omega t+\phi).
\end{aligned}
\label{eq:quadratic-difference-mixing}
\end{equation}

The amplitude \(B_{1}\) in Eq.~\eqref{eq:quadratic-difference-mixing} depends on \(J_{y}\), the linear response
vectors \(\mathbf{H}_{1,2}\), and the nonlinear derivatives of the HH
vector field. Schematically,

\begin{equation}
\begin{aligned}
B_{1}e^{i\phi} &\propto \mathbf{w}^{\dagger}
Q\bigl(\mathbf{H}_{2},\overline{\mathbf{H}_{1}}\bigr),
\end{aligned}
\label{eq:quadratic-mixing-projection}
\end{equation}

where \(\mathbf{w}\) projects the full state dynamics onto the
dynamically relevant excitable direction.

The averaged difference-frequency term in Eq.~\eqref{eq:quadratic-difference-mixing} provides an analytical bridge
between the two-tone stimulus and the slow modulation of neuronal
excitability: the HH nonlinearity itself performs difference-frequency
mixing.

After averaging over the fast carrier, the local dynamics near threshold
can be represented generically by a fast excitable variable \(u\), one
or more recovery variables \(r\), and the slow interference phase
\(\psi\):

\begin{equation}
\begin{aligned}
\dot{u} &= F(u,r;\mu) \\
&\quad + B(J_y,\omega,\Delta\omega)\cos\psi + \cdots,
\end{aligned}
\label{eq:reduced-fast-dynamics}
\end{equation}

\begin{equation}
\begin{aligned}
\dot{r} &= \varepsilon G(u,r),
\end{aligned}
\label{eq:reduced-recovery-dynamics}
\end{equation}

\begin{equation}
\begin{aligned}
\dot{\psi} &= \Delta\omega.
\end{aligned}
\label{eq:reduced-beat-phase}
\end{equation}

Close to a generic threshold, one may write a local normal-form
representation schematically as

\begin{equation}
\begin{aligned}
\dot{u} &= \mu_{\mathrm{eff}} + au^{2} - br + B\cos\psi + \cdots,
\end{aligned}
\label{eq:local-excitability-normal-form}
\end{equation}

with $\mu_{\mathrm{eff}} = \mu_{0} + \mu_{DC}\left( J_{y} \right) + B_{0}$. Equation~\eqref{eq:local-excitability-normal-form} should be interpreted as a local reduced description, not as a
replacement for the HH equations. It shows how the slow phase
periodically moves the neuron towards and away from the excitation
threshold, making the response highly sensitive to detuning. \\

\textbf{\textit{Event-based map}}--- A conventional oscillator phase equation assumes a stable autonomous
limit cycle with a well-defined phase response curve. In the present HH
regime, however, the unforced neuron is excitable and rests at a stable
equilibrium. Therefore, it would be unjustified to assume from
the outset an autonomous phase equation of the form

\begin{equation}
\begin{aligned}
\dot{\theta} &= \omega_{0} + K\sin(\cdots).
\end{aligned}
\label{eq:autonomous-phase-example}
\end{equation}

A more defensible construction uses the spike events generated by the
forced system itself. This creates a Poincaré-type event map without
assuming that the unforced neuron possesses a self-sustained limit
cycle. \\

\textbf{\textit{Definition of spike events and sampled beat
phase}}--- Let $t_{0},t_{1},t_{2},\ldots$ be successive upward crossings of the chosen threshold. At each spike,
sample the slow beat phase:

\begin{equation}
\begin{aligned}
\psi_{k} &= \Delta\omega t_k \pmod{2\pi}.
\end{aligned}
\label{eq:sampled-beat-phase}
\end{equation}

Normalise it to the unit circle,

\begin{equation}
\begin{aligned}
x_k &= \frac{\psi_k}{2\pi} \pmod{1}.
\end{aligned}
\label{eq:normalized-event-phase}
\end{equation}

Define the interspike interval

\begin{equation}
\begin{aligned}
\tau_{k} &= t_{k + 1} - t_{k}.
\end{aligned}
\label{eq:interspike-interval}
\end{equation}

If the recovery state after transients is slaved sufficiently strongly
to the beat phase, the next interspike interval can be represented as a
periodic function

\begin{equation}
\begin{aligned}
\tau_k &= \tau(\psi_k),\\
\tau(\psi+2\pi) &= \tau(\psi).
\end{aligned}
\label{eq:phase-dependent-isi}
\end{equation}

The validity of the one-dimensional relation in Eq.~\eqref{eq:phase-dependent-isi} is not automatic. It must
be checked numerically by plotting \(\tau_{k}\) against \(\psi_{k}\). A
single-valued curve supports the one-dimensional reduction; multiple
branches or a broad cloud would indicate that an additional recovery
coordinate is required. \\

\textbf{\textit{Fourier representation of the phase-dependent return
time}}--- Any sufficiently regular \(2\pi\)-periodic return-time function can be
expanded as

\begin{equation}
\begin{aligned}
\tau(\psi) &= \bar{\tau}
+\sum_{n=1}^{\infty}\bigl[a_n\cos(n\psi)\\
&\qquad\qquad\qquad +b_n\sin(n\psi)\bigr].
\end{aligned}
\label{eq:isi-fourier-series}
\end{equation}

Because the averaged nonlinear forcing is expected to be dominated by
the first difference-frequency harmonic, the leading approximation is

\begin{equation}
\begin{aligned}
\tau(\psi) \simeq \bar{\tau} + a_{1}\cos\psi + b_{1}\sin\psi.
\end{aligned}
\label{eq:isi-first-harmonic}
\end{equation}

Equivalently,

\begin{equation}
\begin{aligned}
\tau(\psi) &\simeq \bar{\tau}+R\sin(\psi+\phi_0),\\
R &= \sqrt{a_1^2+b_1^2}.
\end{aligned}
\label{eq:isi-amplitude-phase}
\end{equation}

This relation is directly testable. From the full HH simulations one
should plot \(t_{k + 1} - t_{k}\) versus
\(\Delta\omega t_{k} \pmod{2\pi}\) and fit the result to Eq.~\eqref{eq:isi-amplitude-phase}. A dominant first harmonic is
quantitative evidence that a sine circle map is an appropriate
leading-order reduction. \\

\textbf{\textit{Derivation of the circle
map}}--- Because \(\psi(t) = \Delta\omega t\), the relative phase at the next
spike is

\begin{equation}
\begin{aligned}
\psi_{k+1} &= \Delta\omega t_{k+1}\\
&= \Delta\omega(t_k+\tau_k).
\end{aligned}
\label{eq:next-spike-phase}
\end{equation}

Hence

\begin{equation}
\begin{aligned}
\psi_{k + 1} &= \psi_{k} + \Delta\omega\tau\left( \psi_{k} \right).
\end{aligned}
\label{eq:phase-return-equation}
\end{equation}

Using the first-harmonic approximation in Eq.~\eqref{eq:isi-amplitude-phase},

\begin{equation}
\begin{aligned}
\psi_{k+1} &= \psi_k+\Delta\omega\bar{\tau}\\
&\quad+\Delta\omega R\sin(\psi_k+\phi_0).
\end{aligned}
\label{eq:first-harmonic-phase-return}
\end{equation}

A constant shift of the phase origin absorbs \(\phi_{0}\). Using the
normalised phase \(x_{k}\) of Eq.~\eqref{eq:normalized-event-phase}, the
phase update can be written in the standard sine-circle-map form

\begin{equation}
\begin{aligned}
x_{k+1} &= x_k+\Omega \\
&\quad-\frac{K}{2\pi}\sin(2\pi x_k)\pmod{1},
\end{aligned}
\label{eq:sine-circle-map}
\end{equation}

where

\begin{equation}
\begin{aligned}
\Omega &= \frac{\Delta\omega\bar{\tau}}{2\pi},\\
K &= |\Delta\omega|R.
\end{aligned}
\label{eq:circle-map-parameters}
\end{equation}

The sign of the sinusoidal term can be changed by a phase shift, so the
choice of the conventional minus sign in this standard form does not
affect the qualitative mode-locking structure.

The physical interpretation is important:

\begin{itemize}
\item
  \(\Delta\omega\) primarily controls the bare rotation parameter
  \(\Omega\).
\item
  \(J_{y}\) changes the nonlinear excitability and therefore the
  modulation amplitude \(R\), which changes the effective coupling
  \(K\).
\item
  The physical parameter plane \(\left( \Delta\omega,J_{y} \right)\)
  therefore maps naturally onto the circle-map plane \((\Omega,K)\).
\end{itemize}

Equation~\eqref{eq:sine-circle-map} is the central reduced description used below for the mode-locking analysis. \\

\textbf{\textit{Rotation
number}}--- For mode-locking analysis one must not discard the accumulated phase by
applying modulo one at every step. Define the lift

\begin{equation}
\begin{aligned}
X_{k+1} &= X_k+\Omega\\
&\quad-\frac{K}{2\pi}\sin(2\pi X_k),\\
X_k &\in \mathbb{R}.
\end{aligned}
\label{eq:lifted-circle-map}
\end{equation}

The circle variable is recovered through

\begin{equation}
\begin{aligned}
x_k &= X_k\pmod{1}.
\end{aligned}
\label{eq:modulo-circle-state}
\end{equation}

The winding or rotation number is

\begin{equation}
\begin{aligned}
\rho &= \lim_{N \rightarrow \infty}\frac{X_{N} - X_{0}}{N}.
\end{aligned}
\label{eq:lift-rotation-number}
\end{equation}

Equivalently, in radians,

\begin{equation}
\begin{aligned}
\rho &= \lim_{N\to\infty}
\frac{\Psi_N-\Psi_0}{2\pi N},
\end{aligned}
\label{eq:unwrapped-rotation-number}
\end{equation}

where \(\Psi_{N}\) denotes the unwrapped beat phase sampled at
successive spikes.

When \(K = 0\),

\begin{equation}
\begin{aligned}
X_{k + 1} &= X_{k} + \Omega,
\end{aligned}
\label{eq:zero-coupling-lift}
\end{equation}

and therefore

\begin{equation}
\begin{aligned}
\rho &= \Omega.
\end{aligned}
\label{eq:zero-coupling-rotation}
\end{equation}

For nonzero coupling, the map can lock over finite parameter intervals
to rational values

\begin{equation}
\begin{aligned}
\rho &= \frac{p}{q}.
\end{aligned}
\label{eq:rational-rotation}
\end{equation}

The rational locking condition in Eq.~\eqref{eq:rational-rotation} implies

\begin{equation}
\begin{aligned}
X_{k + q} &= X_{k} + p,
\end{aligned}
\label{eq:lift-periodic-orbit}
\end{equation}

or, in angular variables,

\begin{equation}
\begin{aligned}
\Psi_{k + q} &= \Psi_{k} + 2\pi p.
\end{aligned}
\label{eq:angle-periodic-orbit}
\end{equation}

Thus, after \(q\) spike events, the beat phase has advanced by \(p\)
complete cycles. \\

\textbf{\textit{Relation between the rotation number and the physical
spike/beat frequency
ratio}}--- Because the map is sampled once per spike, the interpretation of \(p/q\)
requires care. If \(q\) spikes occur while the beat phase advances by
\(p\) cycles, then

\begin{equation}
\begin{aligned}
q\ \text{spikes} &\quad\leftrightarrow\quad p\ \text{beat cycles}.
\end{aligned}
\label{eq:spike-beat-correspondence}
\end{equation}

Therefore

\begin{equation}
\begin{aligned}
\frac{f_{\mathrm{out}}}{f_b} &= \frac{q}{p}=\frac{1}{\rho},
\end{aligned}
\label{eq:spike-to-beat-ratio}
\end{equation}

with

\begin{equation}
\begin{aligned}
f_{b} &= \frac{|\Delta\omega|}{2\pi}.
\end{aligned}
\label{eq:beat-frequency}
\end{equation}

A convenient quantity to report directly from HH spike times is

\begin{equation}
\begin{aligned}
W &\equiv \frac{f_{\mathrm{out}}}{f_b}.
\end{aligned}
\label{eq:spike-beat-ratio-definition}
\end{equation}

If \(N\) interspike intervals span the time \(t_{N} - t_{0}\), then

\begin{equation}
\begin{aligned}
f_{\mathrm{out}} &= \lim_{N\to\infty}
\frac{N}{t_N-t_0},
\end{aligned}
\label{eq:output-firing-frequency}
\end{equation}

so

\begin{equation}
\begin{aligned}
W &= \frac{2\pi}{|\Delta\omega|}
\lim_{N\to\infty}\frac{N}{t_N-t_0}.
\end{aligned}
\label{eq:spike-beat-ratio-estimator}
\end{equation}

The quantity \(\left( t_{N} - t_{0} \right)/N\) is the mean interspike
\textbf{period}, not the firing frequency.

For \(\Delta\omega = 0\), \(f_{b} = 0\) and the spike/beat ratio is
undefined. The zero-detuning case should therefore be treated separately
rather than assigned an artificial winding number. \\

\textbf{\textit{Rational locking and Arnold
tongues}}--- For a rational rotation number $\rho = \frac{p}{q}$, a \(q\)-periodic orbit of the lifted map satisfies

\begin{equation}
\begin{aligned}
F^{q}\left( X^{*} \right) &= X^{*} + p.
\end{aligned}
\label{eq:q-cycle-periodicity}
\end{equation}

A stable locked orbit additionally satisfies

\begin{equation}
\begin{aligned}
\left| \frac{dF^{q}}{dX}\left( X^{*} \right) \right| < 1.
\end{aligned}
\label{eq:locked-orbit-stability}
\end{equation}

The boundary of an Arnold tongue occurs at a saddle-node bifurcation of
the periodic orbit. Therefore the boundary conditions are

\begin{equation}
\begin{aligned}
F^{q}(X^{*}) &= X^{*}+p,
\end{aligned}
\label{eq:arnold-tongue-periodic-condition}
\end{equation}

and

\begin{equation}
\begin{aligned}
\frac{dF^{q}}{dX}(X^{*}) &= 1.
\end{aligned}
\label{eq:arnold-tongue-tangency-condition}
\end{equation}

Equations~\eqref{eq:arnold-tongue-periodic-condition} and~\eqref{eq:arnold-tongue-tangency-condition} provide a precise analytical definition of a tongue boundary, stronger than identifying a visual lobe or dip in the
threshold curve.

For the sine circle map,

\begin{equation}
\begin{aligned}
F'(X) &= 1-K\cos(2\pi X),
\end{aligned}
\label{eq:circle-map-derivative}
\end{equation}

so

\begin{equation}
\begin{aligned}
\frac{dF^q}{dX} &= \prod_{j=0}^{q-1}
\bigl[1-K\cos(2\pi X_j)\bigr].
\end{aligned}
\label{eq:q-cycle-multiplier}
\end{equation}

Thus a general \(p/q\) boundary satisfies the periodic-orbit and
saddle-node conditions above, or equivalently the periodic-orbit
condition together with

\begin{equation}
\begin{aligned}
\prod_{j=0}^{q-1}\bigl[1-K\cos(2\pi X_j)\bigr] &= 1.
\end{aligned}
\label{eq:arnold-tongue-product-condition}
\end{equation}

Equations~\eqref{eq:arnold-tongue-periodic-condition} and~\eqref{eq:arnold-tongue-product-condition} can be solved numerically for the reduced map and
compared with the full HH locking diagram. \\

\textbf{\textit{Analytical boundary of the simplest period-one
tongue}}--- For a period-one locked state with integer rotation \(\rho = p\), the
lifted map obeys $F\left( X^{*} \right) = X^{*} + p$.

Using the lifted sine circle map,

\begin{equation}
\begin{aligned}
\Omega - \frac{K}{2\pi}\sin\left( 2\pi X^{*} \right) &= p.
\end{aligned}
\label{eq:integer-locking-equation}
\end{equation}

Therefore

\begin{equation}
\begin{aligned}
\sin\left( 2\pi X^{*} \right) &= \frac{2\pi(\Omega - p)}{K}.
\end{aligned}
\label{eq:integer-locking-sine}
\end{equation}

A solution exists only if the magnitude of the right-hand side is no
larger than one:

\begin{equation}
\begin{aligned}
|\Omega - p| \leq \frac{|K|}{2\pi}.
\end{aligned}
\label{eq:period-one-locking-existence}
\end{equation}

Hence the two boundaries are

\begin{equation}
\begin{aligned}
\Omega_p^{\pm} &= p\pm\frac{K}{2\pi},
\end{aligned}
\label{eq:period-one-tongue-edges}
\end{equation}

and the tongue width in \(\Omega\) is

\begin{equation}
\begin{aligned}
\Delta\Omega_{p} &= \frac{|K|}{\pi}.
\end{aligned}
\label{eq:period-one-tongue-width}
\end{equation}

Equation~\eqref{eq:period-one-tongue-width} explicitly shows that stronger phase-dependent coupling broadens
the locking region.

Using the lowest-order relations
\(\Omega = \Delta\omega\bar{\tau}/(2\pi)\) and
\(K \simeq \Delta\omega R\) for positive detuning, the existence condition becomes

\begin{equation}
\begin{aligned}
\left|\frac{\Delta\omega\bar{\tau}}{2\pi}-p\right|
&\le \frac{\Delta\omega R}{2\pi}.
\end{aligned}
\label{eq:period-one-existence-detuning}
\end{equation}

At the boundaries,

\begin{equation}
\begin{aligned}
\Delta\omega\bar{\tau}-2\pi p &= \pm\Delta\omega R,
\end{aligned}
\label{eq:detuning-edge-condition}
\end{equation}

which gives the approximate detuning locations

\begin{equation}
\begin{aligned}
\Delta\omega_p^{\pm} &\simeq
\frac{2\pi p}{\bar{\tau}\mp R}.
\end{aligned}
\label{eq:approximate-tongue-detunings}
\end{equation}

In the full HH model, \(\bar{\tau}\) and \(R\) can themselves depend on
\(J_{y}\) and \(\Delta\omega\), so this detuning estimate should be
regarded as a leading-order prediction, not an exact boundary formula. \\

\textbf{\textit{Farey hierarchy and devil's-staircase
organisation}}--- For fixed coupling, the circle map produces intervals in which the
rotation number remains locked at rational values. Thus the graph of
\(\rho\) versus \(\Omega\) develops plateaus. The organising principle
is the Farey hierarchy.

Suppose two prominent neighbouring plateaus have rotation numbers

\begin{equation}
\begin{aligned}
\frac{p_1}{q_1} \qquad\text{and}\qquad \frac{p_2}{q_2}.
\end{aligned}
\label{eq:farey-neighbor-fractions}
\end{equation}

They are Farey neighbours when

\begin{equation}
\begin{aligned}
\left|p_1q_2-p_2q_1\right| &= 1.
\end{aligned}
\label{eq:farey-neighbor-criterion}
\end{equation}

The principal rational resonance expected between them is the Farey mediant

\begin{equation}
\begin{aligned}
\frac{p_1+p_2}{q_1+q_2}.
\end{aligned}
\label{eq:farey-mediant}
\end{equation}

For example, $\frac{1}{3}$ and $\frac{1}{2}$ lead to $\frac{1 + 1}{3 + 2} = \frac{2}{5}$. Similarly, $\frac{1}{2}$ and $\frac{2}{3}$ lead to $\frac{3}{5}$. Therefore, a convincing demonstration of Farey organisation would
identify low-order rational plateaus and then show that intermediate
locking states appear in the expected mediant order.

The term \textbf{devil's staircase} should however be used conservatively.
Rational mode-locking plateaus and Farey ordering can be tested
directly. A claim of a complete devil's staircase requires stronger
evidence about the coverage of the control-parameter interval and the
effective map regime. For the canonical sine circle map, the derivative
is given by Eq.~\eqref{eq:circle-map-derivative}.
For \(K < 1\) the map is monotone and invertible; at \(K = 1\) it
reaches the critical onset of noninvertibility; for \(K > 1\) it is
noninvertible. The precise strength of the claim should therefore follow
the measured effective map rather than be assumed in advance. \\

\textbf{\textit{Why does increasing the observation time reveal
progressively finer
structures?}}--- Figure~\ref{fig:farey} shows that increasing the maximum integration time
produces progressively finer tongue-like structures. The circle-map framework gives a specific explanation.

A rationally locked state with $\rho = \frac{p}{q}$ satisfies $F^{q}(X) = X + p$. Therefore at least \(q\) spike events are needed to see one full period
of the event map. In practice several periods are needed for reliable
classification. If \(M\) repetitions are required, the minimum
observation time scales approximately as

\begin{equation}
\begin{aligned}
T_{\mathrm{obs}} &\gtrsim Mq\bar{\tau}.
\end{aligned}
\label{eq:minimum-observation-time}
\end{equation}

Because \(q\) spikes correspond to \(p\) beat periods,

\begin{equation}
\begin{aligned}
T_{\mathrm{obs}} &\gtrsim MpT_b.
\end{aligned}
\label{eq:beat-period-observation-time}
\end{equation}

Consequently:

\begin{itemize}
	\item short observation windows preferentially reveal low-order rational states;
	\item longer observation windows can resolve higher-order resonances;
	\item higher-order Farey descendants typically occupy narrower parameter intervals;
	\item progressively finer apparent structure can therefore emerge as the integration time is increased.
\end{itemize}

Equations~\eqref{eq:minimum-observation-time} and~\eqref{eq:beat-period-observation-time} provide a more specific explanation than describing the refinement only as visual self-similarity. They also provide a convergence test: if
a feature corresponds to true asymptotic mode locking, its measured rational rotation number should stabilise as the observation window is
increased.

\section{Conclusion}

In this work, we demonstrate that temporal interference stimulation fundamentally reorganises the excitability landscape of the Hodgkin--Huxley neuron, giving rise to the phenomenon of ``multitongue frequency fractals''. Unlike the smooth single-frequency response, dual-frequency drive fragments the excitability profile into hierarchically structured lobes and dips that refine with increasing observation time, revealing clear self-similar structure quantified by a box-counting-style scaling exponent $D = 1.36 \pm 0.12$. This structure arises from the coexistence of non-cascaded high-harmonic generation, cascaded high-harmonic generation and cascaded subharmonic generation, and from the sharp switching between these regimes as detuning varies. The phenomenon is resonantly organised, becoming most pronounced when the carrier frequency aligns with intrinsic ionic timescales, thereby maximising nonlinear envelope interactions. The proliferation of resolved tongues with $T_{\max}$ follows a Farey-tree hierarchy and is quantitatively consistent with a bounded box-counting-style scaling law, demonstrating that multitongue frequency fractals constitute a robust, quantifiable synchronisation phenomenon in Hodgkin--Huxley dynamics under temporal interference stimulation.

\section*{References}

\bibliography{apssamp}

@inproceedings{plovie2022influence,
  author = {Plovie, Tom and Schoeters, Ruben and Tarnaud, Thomas and Martens, Luc and Joseph, Wout and Tanghe, Emmeric},
  title = {Influence of temporal interference stimulation parameters on point neuron excitability},
  booktitle = {2022 44th Annual International Conference of the {IEEE} Engineering in Medicine and Biology Society ({EMBC})},
  pages = {2365--2368},
  year = {2022},
  month = {July},
  publisher = {IEEE},
  organization = {IEEE},
  doi = {10.1109/EMBC48229.2022.9871641},
}

@inproceedings{plovie2024contribution,
  author = {Plovie, Tom and Schoeters, Ruben and Tarnaud, Thomas and Joseph, Wout and Tanghe, Emmeric},
  title = {Contribution of ion channel nonlinearities to the mechanism of temporal interference stimulation},
  booktitle = {BioEM 2024: The 3rd Annual Conference of {BioEM}, Full Abstract Book},
  pages = {200},
  year = {2024},
  month = {June},
  address = {Crete, Greece},
  note = {Abstract OS03-03; eight-page full abstract},
}

@article{plovie2025nonlinearities,
  author = {Plovie, Tom and Schoeters, Ruben and Tarnaud, Thomas and Joseph, Wout and Tanghe, Emmeric},
  title = {Nonlinearities and timescales in neural models of temporal interference stimulation},
  journal = {Bioelectromagnetics},
  volume = {46},
  number = {1},
  pages = {e22522},
  year = {2025},
  doi = {10.1002/bem.22522},
}

@article{parmananda2002resonant,
  author = {Parmananda, P. and Mena, Claudia H. and Baier, Gerold},
  title = {Resonant forcing of a silent {Hodgkin--Huxley} neuron},
  journal = {Physical Review E},
  volume = {66},
  number = {4},
  pages = {047202},
  year = {2002},
  doi = {10.1103/PhysRevE.66.047202},
}

@article{parmananda2001resonance,
  author = {Parmananda, P. and Mahara, Hitoshi and Amemiya, Takashi and Yamaguchi, Tomohiko},
  title = {Resonance-induced pacemakers: A new class of organizing centers for wave propagation in excitable media},
  journal = {Physical Review Letters},
  volume = {87},
  number = {23},
  pages = {238302},
  year = {2001},
  doi = {10.1103/PhysRevLett.87.238302},
}

@article{Hodgkin1952,
  author = {Hodgkin, Alan L. and Huxley, Andrew F.},
  title = {A quantitative description of membrane current and its application to conduction and excitation in nerve},
  journal = {The Journal of Physiology},
  volume = {117},
  number = {4},
  pages = {500--544},
  year = {1952},
  doi = {10.1113/jphysiol.1952.sp004764},
}

@article{FitzHugh1961,
  author = {FitzHugh, Richard},
  title = {Impulses and physiological states in theoretical models of nerve membrane},
  journal = {Biophysical Journal},
  volume = {1},
  number = {6},
  pages = {445--466},
  year = {1961},
  doi = {10.1016/S0006-3495(61)86902-6},
}

@article{Mirzakhalili2020,
  author = {Mirzakhalili, Ehsan and Barra, Beatrice and Capogrosso, Marco and Lempka, Scott F.},
  title = {Biophysics of temporal interference stimulation},
  journal = {Cell Systems},
  volume = {11},
  number = {6},
  pages = {557--572.e5},
  year = {2020},
  month = {December},
  doi = {10.1016/j.cels.2020.10.004},
}

@article{Esmaeilpour2021,
  author = {Esmaeilpour, Zeinab and Kronberg, Greg and Reato, Davide and Parra, Lucas C. and Bikson, Marom},
  title = {Temporal interference stimulation targets deep brain regions by modulating neural oscillations},
  journal = {Brain Stimulation},
  volume = {14},
  number = {1},
  pages = {55--65},
  year = {2021},
  doi = {10.1016/j.brs.2020.11.007},
}

@article{Meron1992,
  author = {Meron, Ehud},
  title = {Pattern formation in excitable media},
  journal = {Physics Reports},
  volume = {218},
  number = {1},
  pages = {1--66},
  year = {1992},
  doi = {10.1016/0370-1573(92)90098-K},
}

@book{Izhikevich2007,
  author = {Izhikevich, Eugene M.},
  title = {Dynamical Systems in Neuroscience: The Geometry of Excitability and Bursting},
  year = {2007},
  publisher = {MIT Press},
  address = {Cambridge, MA},
  isbn = {978-0-262-09043-8},
  doi = {10.7551/mitpress/2526.001.0001},
}

@book{Pikovsky2001,
  author = {Pikovsky, Arkady and Rosenblum, Michael and Kurths, J{\"u}rgen},
  title = {Synchronization: A Universal Concept in Nonlinear Sciences},
  number = {12},
  year = {2001},
  publisher = {Cambridge University Press},
  address = {Cambridge},
  series = {Cambridge Nonlinear Science Series},
  isbn = {978-0-521-59285-7},
  doi = {10.1017/CBO9780511755743},
}

@article{Grossman2017,
  author = {Grossman, Nir and Bono, David and Dedic, Nina and Kodandaramaiah, Suhasa B. and Rudenko, Andrii and Suk, Ho-Jun and Cassar{\`a}, Antonino M. and Neufeld, Esra and Kuster, Niels and Tsai, Li-Huei and Pascual-Leone, Alvaro and Boyden, Edward S.},
  title = {Noninvasive deep brain stimulation via temporally interfering electric fields},
  journal = {Cell},
  volume = {169},
  number = {6},
  pages = {1029--1041.e16},
  year = {2017},
  month = {June},
  doi = {10.1016/j.cell.2017.05.024},
}

@article{yuan2016theoretical,
  author = {Yuan, Yi and Chen, Yudong and Li, Xiaoli},
  title = {Theoretical analysis of transcranial magneto-acoustical stimulation with a {Hodgkin--Huxley} neuron model},
  journal = {Frontiers in Computational Neuroscience},
  volume = {10},
  pages = {35},
  year = {2016},
  doi = {10.3389/fncom.2016.00035},
}

@article{johnson2017spike,
  author = {Johnson, Melissa G. and Chartier, Sylvain},
  title = {Spike neural models (part {I}): The {Hodgkin--Huxley} model},
  journal = {The Quantitative Methods for Psychology},
  volume = {13},
  number = {2},
  pages = {105--119},
  year = {2017},
  doi = {10.20982/tqmp.13.2.p105},
}

@article{levi2018digital,
  author = {Levi, Timoth{\'e}e and Khoyratee, Farad and Sa{\"i}ghi, Sylvain and Ikeuchi, Yoshiho},
  title = {Digital implementation of the {Hodgkin--Huxley} neuron model for neurological disease studies},
  journal = {Artificial Life and Robotics},
  volume = {23},
  number = {1},
  pages = {10--14},
  year = {2018},
  publisher = {Springer},
  doi = {10.1007/s10015-017-0397-7},
}

@article{giannari2022model,
  author = {Giannari, Anastasia G. and Astolfi, Alessandro},
  title = {Model design for networks of heterogeneous {Hodgkin--Huxley} neurons},
  journal = {Neurocomputing},
  volume = {496},
  pages = {147--157},
  year = {2022},
  doi = {10.1016/j.neucom.2022.04.115},
}

@inproceedings{kumar2016design,
  author = {Kumar, Jayanth and Kumar, Juneeth and Murali, Shanmukha and Bhakthavatchalu, Ramesh},
  title = {Design and implementation of {Izhikevich}, {Hodgkin}, and {Huxley} spiking neuron models and their comparison},
  booktitle = {2016 International Conference on Advanced Communication Control and Computing Technologies ({ICACCCT})},
  pages = {111--116},
  year = {2016},
  publisher = {IEEE},
  organization = {IEEE},
  doi = {10.1109/ICACCCT.2016.7831611},
}

@incollection{nelson1998hodgkin,
  author = {Nelson, Mark E. and Rinzel, John},
  title = {The {Hodgkin--Huxley} model},
  booktitle = {The Book of {GENESIS}: Exploring Realistic Neural Models with the {GEneral NEural SImulation System}},
  editor = {Bower, James M. and Beeman, David},
  pages = {29--49},
  year = {1998},
  publisher = {Springer},
  address = {New York},
  edition = {2},
  doi = {10.1007/978-1-4612-1634-6_4},
}

@article{rutherford2020analog,
  author = {Rutherford, George H. and Mobille, Zach D. and Brandt-Trainer, Jordan and Follmann, Rosangela and Rosa, Jr., Epaminondas},
  title = {Analog implementation of a {Hodgkin--Huxley} model neuron},
  journal = {American Journal of Physics},
  volume = {88},
  number = {11},
  pages = {918--923},
  year = {2020},
  doi = {10.1119/10.0001072},
}

@article{aihara_periodic_1984,
  author = {Aihara, Kazuyuki and Matsumoto, Gen and Ikegaya, Yuhji},
  title = {Periodic and non-periodic responses of a periodically forced {Hodgkin--Huxley} oscillator},
  journal = {Journal of Theoretical Biology},
  volume = {109},
  number = {2},
  pages = {249--269},
  year = {1984},
  month = {July},
  doi = {10.1016/S0022-5193(84)80005-3},
  url = {https://linkinghub.elsevier.com/retrieve/pii/S0022519384800053},
}

@incollection{degn_forced_1987,
  author = {Aihara, K. and Matsumoto, G.},
  title = {Forced oscillations and routes to chaos in the {Hodgkin--Huxley} axons and squid giant axons},
  booktitle = {Chaos in Biological Systems},
  editor = {Degn, H. and Holden, A. V. and Olsen, L. F.},
  pages = {121--131},
  year = {1987},
  publisher = {Springer US},
  address = {Boston, MA},
  isbn = {978-1-4757-9633-9 978-1-4757-9631-5},
  doi = {10.1007/978-1-4757-9631-5_15},
  url = {http://link.springer.com/10.1007/978-1-4757-9631-5_15},
}

@article{che_phase-locking_2009,
  author = {Che, Yan-Qiu and Wang, Jiang and Si, Wen-Jie and Fei, Xiang-Yang},
  title = {Phase-locking and chaos in a silent {Hodgkin--Huxley} neuron exposed to a sinusoidal electric field},
  journal = {Chaos, Solitons \& Fractals},
  volume = {39},
  number = {1},
  pages = {454--462},
  year = {2009},
  month = {January},
  doi = {10.1016/j.chaos.2007.03.008},
  url = {https://linkinghub.elsevier.com/retrieve/pii/S0960077907002780},
}

@article{lee_bifurcation_2006,
  author = {Lee, Sang-Gui and Kim, Seunghwan},
  title = {Bifurcation analysis of mode-locking structure in a {Hodgkin--Huxley} neuron under sinusoidal current},
  journal = {Physical Review E},
  volume = {73},
  number = {4},
  pages = {041924},
  year = {2006},
  month = {April},
  doi = {10.1103/PhysRevE.73.041924},
  url = {https://link.aps.org/doi/10.1103/PhysRevE.73.041924},
}

@article{abhay_high-order_2025,
  author = {{Abhay} and Dar, Gaurav},
  title = {High-order synchronization in identical neurons with asymmetric pulse coupling},
  journal = {Physical Review E},
  volume = {111},
  number = {3},
  pages = {034202},
  year = {2025},
  month = {March},
  doi = {10.1103/PhysRevE.111.034202},
  url = {https://link.aps.org/doi/10.1103/PhysRevE.111.034202},
}

@article{borkowski_bistability_2011,
  author = {Borkowski, L. S.},
  title = {Bistability and resonance in the periodically stimulated {Hodgkin--Huxley} model with noise},
  journal = {Physical Review E},
  volume = {83},
  number = {5},
  pages = {051901},
  year = {2011},
  month = {May},
  doi = {10.1103/PhysRevE.83.051901},
  url = {https://link.aps.org/doi/10.1103/PhysRevE.83.051901},
}

@article{allen_arithmetic_1983,
  author = {Allen, Tracy},
  title = {On the arithmetic of phase locking: Coupled neurons as a lattice on {$R^2$}},
  journal = {Physica D: Nonlinear Phenomena},
  volume = {6},
  number = {3},
  pages = {305--320},
  year = {1983},
  month = {April},
  doi = {10.1016/0167-2789(83)90014-3},
  url = {https://linkinghub.elsevier.com/retrieve/pii/0167278983900143},
}

@article{felicio_arnold_2015,
  author = {Felicio, Carolini C. and Rech, Paulo C.},
  title = {Arnold tongues and the devil's staircase in a discrete-time {Hindmarsh--Rose} neuron model},
  journal = {Physics Letters A},
  volume = {379},
  number = {43--44},
  pages = {2845--2847},
  year = {2015},
  month = {November},
  doi = {10.1016/j.physleta.2015.09.021},
  url = {https://linkinghub.elsevier.com/retrieve/pii/S0375960115008087},
}

@article{borkowski_response_2009,
  author = {Borkowski, L. S.},
  title = {Response of a {Hodgkin--Huxley} neuron to a high-frequency input},
  journal = {Physical Review E},
  volume = {80},
  number = {5},
  pages = {051914},
  year = {2009},
  month = {November},
  doi = {10.1103/PhysRevE.80.051914},
  url = {https://link.aps.org/doi/10.1103/PhysRevE.80.051914},
}

@article{guevara_glass_1982,
	title = {Phase locking, period doubling bifurcations and chaos in a mathematical model of a periodically driven oscillator: {A} theory for the entrainment of biological oscillators and the generation of cardiac dysrhythmias},
	volume = {14},
	issn = {0303-6812, 1432-1416},
	shorttitle = {Phase locking, period doubling bifurcations and chaos in a mathematical model of a periodically driven oscillator},
	url = {https://link.springer.com/article/10.1007/BF02154750},
	doi = {10.1007/BF02154750},
	number = {1},
	urldate = {2026-07-12},
	journal = {Journal of Mathematical Biology},
	author = {Guevara, Michael R. and Glass, Leon},
	month = mar,
	year = {1982},
	pages = {1--23},
}

@article{perez_glass_1982,
	title = {Bistability, period doubling bifurcations and chaos in a periodically forced oscillator},
	volume = {90},
	issn = {0375-9601},
	url = {https://www.sciencedirect.com/science/article/pii/0375960182903917},
	doi = {10.1016/0375-9601(82)90391-7},
	number = {9},
	urldate = {2026-07-12},
	journal = {Physics Letters A},
	author = {Perez, Rafael and Glass, Leon},
	month = aug,
	year = {1982},
	pages = {441--443},
}

@article{grebogi_ott_yorke_1985,
	title = {Attractors on an {N}-torus: {Quasiperiodicity} versus chaos},
	volume = {15},
	issn = {0167-2789},
	shorttitle = {Attractors on an {N}-torus},
	url = {https://www.sciencedirect.com/science/article/pii/016727898580004X},
	doi = {10.1016/S0167-2789(85)80004-X},
	number = {3},
	urldate = {2026-07-12},
	journal = {Physica D: Nonlinear Phenomena},
	author = {Grebogi, Celso and Ott, Edward and Yorke, James A.},
	month = apr,
	year = {1985},
	pages = {354--373},
}

@article{lim_kim_kim_2009,
	title = {Strange nonchaotic responses of the quasiperiodically forced {Morris}-{Lecar} neuron},
	volume = {121},
	issn = {0033-068X},
	url = {https://academic.oup.com/ptp/article/121/4/671/1854701},
	doi = {10.1143/PTP.121.671},
	number = {4},
	urldate = {2026-07-12},
	journal = {Progress of Theoretical Physics},
	author = {Lim, Woochang and Kim, Sang-Yoon and Kim, Youngtae},
	month = apr,
	year = {2009},
	pages = {671--683},
}

\end{document}